\documentclass[11pt]{article}   
\pdfoutput=1

\usepackage{amsthm,amsmath,amssymb,bbold,xcolor,sectsty,latexsym,mathtools,hyperref,mathrsfs,slashed,bm,epsfig}

\usepackage[textwidth=6.5in,top=1.2in,bottom=1.2in]{geometry}

\newcommand{\beq}{\begin{equation}}
\newcommand{\eeq}{\end{equation}}
\newcommand{\bea}{\begin{eqnarray}}
\newcommand{\eea}{\end{eqnarray}}
\newcommand{\amp}{&\!\!\!}

\newcommand{\rr}{R}
\newcommand{\R}{\tilde{R}}
\newcommand{\n}{\tilde{N}}
\newcommand{\Rm}{\tilde{R}_{\star}}
\newcommand{\E}{\tilde{H}}

\newcommand{\Clm}{C_{lm}}
\newcommand{\ma}{m_{\phi}}

\begin{document}

\thispagestyle{empty}
\begin{titlepage}
\nopagebreak

\title{ \begin{center}\bf Scalar Dark Matter Clumps with Angular Momentum \end{center} }

\vfill
\author{Mark P.~Hertzberg$^{}$\footnote{mark.hertzberg@tufts.edu},\,\,\, Enrico D.~Schiappacasse$^{}$\footnote{enrico.schiappacasse@tufts.edu}}

\date{\today}

\maketitle

\begin{center}
	\vspace{-0.7cm}
	{\it  $^{}$Institute of Cosmology, Department of Physics and Astronomy}\\
	{\it  Tufts University, Medford, MA 02155, USA}
	
\end{center}

\bigskip

\begin{abstract}
The behavior of light scalar dark matter has been a subject of much interest recently as it can lead to interesting small scale behavior. In particular, this can lead to the formation of gravitationally bound clumps for these light scalars, including axions. In Ref.~\cite{Schiappacasse:2017ham} we analyzed the behavior of these clumps, assuming spherical symmetry, allowing for both attractive and repulsive self-interactions. There is a maximum allowed mass for the clumps in the case of attractive interactions, and a minimum radius for the clumps in the case of repulsive interactions, which is saturated at large mass. Here we extend this work to include non-spherically symmetric clumps. Since the system tries to re-organize into a BEC ground state, we consider configurations with a conserved non-zero angular momentum, and construct minimum energy configurations at fixed particle number and fixed angular momentum. We find generalizations of the previous spherically symmetric results. In particular, while there is still a maximum mass for the attractive case, its value increases with angular momentum. Also, the minimum radius in the repulsive case is raised to higher radii. We clarify how a recent claim in the literature of an upper bound on angular momentum is due to inaccurate numerics. In a forthcoming paper we shall investigate the possibility of resonance of axion clumps into both visible and hidden sector photons, and analyze how the altered mass and radius from non-zero angular momentum affects the resonance.
\end{abstract}

\end{titlepage}

\setcounter{page}{2}

\tableofcontents

\section{Introduction}

An ongoing challenge of modern physics is to determine the nature of dark matter, which forms the bulk of the matter in the universe. Many astrophysical and cosmological observations, including CMB, large scale structure, galactic rotation curves, and beyond, are consistent with the presence of stable, cold dark matter \cite{Peebles:2013hla}. Furthermore, there is currently little evidence on large scales in favor of modifications of gravity (in fact there is primarily only counter-evidence). On galactic scales, the agreement between cold dark matter and observations is not as clear; for instance, the presence of cores at the centers of galaxies is non-trivial to explain in the dark matter framework \cite{Deng2018}. Nevertheless the idea that there is one (or more) new particles that comprises the dark matter is a perfectly reasonable possibility within the framework of particle physics, especially since the Standard Model of particle physics has various short comings, including the strong CP problem. However, the identity of the dark matter particle/s remains elusive. In the event that dark matter is not directly detected on earth, it is crucial to determine astrophysical signatures that can distinguish between different dark matter candidates.

In this work we continue our investigation from Ref.~\cite{Schiappacasse:2017ham} into the behavior of light bosons, as they may lead to interesting astrophysical consequences. In particular, we are especially interested in the QCD axion, as well as axion-like-particles, and other light scalar dark matter (that in general could have attractive or repulsive self-interactions). Our focus is on the behavior of these particles on small scales. In this case, its initial power spectrum is highly sensitive to the details of the early universe, including inflation. In particular, if these particles were formed only after a phase transition in the post-inflationary era, then causality dictates that its initial spatial distribution will be significantly inhomogeneous from one Hubble patch to the next. In the case of the QCD axion, this means that the axion would be inhomogeneous after the Peccei-Quinn phase transition. Then when the axion acquires a mass after the QCD phase transition, it would begin to evolve and red-shift with these highly inhomogeneous initial conditions on small scales.

An important consequence of these inhomogeneous initial conditions is that the evolution of the axion would be rather non-linear. This large mode-mode coupling can plausibly lead to the formation of a Bose-Einstein condensate (BEC). The reason is the following: we are studying bosonic particles at high occupancy with an (approximately) conserved particle number. These are necessary conditions to form a BEC \cite{Sikivie:2009qn,Erken:2011dz}. The evolution of the field is highly non-linear, and possibly chaotic, but by an appropriate ensemble averaging, the dynamics can be very accurately studied within the context of classical field theory, as shown rigorously in Ref.~\cite{Hertzberg:2016tal}. Since the BEC is a type of ground state, and since it is driven primarily by attractive gravitational interactions, it does not exhibit long-range order. Instead it is a gravitationally bound clump \cite{Guth:2014hsa}; sometimes known as a ``Bose star" (and can organize into so-called ``mini-clusters") \cite{Kolb:1993zz}.

Strictly speaking, a BEC is a ground state configuration, and since there is nothing in the system that will spontaneously break rotational symmetry, one can safely assume that the ground state configuration is spherically symmetric. This is the standard type of Bose star that has been considered in the literature  (see, for example, Refs.~\cite{Chavanis:2011zi, Chavanis:2011zm}), and it was the focus of our previous paper in Ref.~\cite{Schiappacasse:2017ham}. If only gravity is important, then the mass of a gravitationally bound clump and its radius are inversely related $R\propto 1/M$. This is reminiscent of the ground state of the hydrogen atom in which the Bohr radius is inversely proportional to the electron mass. Apart from this inverse relation there is no further constraint on the mass or radius. However, in Ref.~\cite{Schiappacasse:2017ham} we showed that there are further restrictions on mass and/or radius when self-interactions $\Delta V= \lambda\,\phi^4/4!$ are taken into account. In particular, if this interaction is attractive ($\lambda<0$), which is the usual case for an axion, there is an upper limit to the mass and a lower limit to the radius on stable solutions. This is because the attractive self-interaction tries to create a collapse instability when it is dominant. So it must be always sub-dominant, which only occurs for sufficiently dilute clumps. On the other hand, if this interaction is repulsive ($\lambda>0$), which may happen in an ordinary renormalizable scalar field theory, there is no upper limit to the mass, but the radius does saturate towards a minimum value in the large mass regime. This is a type of polytrope with equation of state $P\propto \rho^2$.

In this paper, we would like to generalize these results to non-spherical configurations. In this case we are not in the true ground state of the system, so it may seem unclear as to the relevance of such configurations to a BEC. Hence we will consider non-spherical configurations that are specified by a fixed amount of angular momentum. We will effectively study a different type of BEC; while an ordinary BEC is the true ground state that minimizes energy at fixed particle number, we will study configurations that minimize energy at fixed particle number {\em and} fixed angular momentum. These objects will therefore exhibit rotation.

Rotating BECs have previously been considered in the literature especially in the context of dark matter halos. Rotation curves of halos in the context of ultralight BEC dark matter were studied in Refs.~\cite{Guzman:2013rua, Guzman:2015jba, Zhang:2018okg}. In this context, the rotation of the galatic halo may cause the formation of vortices from superfluidic behavour of the BEC, as was first discussed in Ref.~\cite{Silverman:2001} (for more recent studies, see, for example, Refs.~\cite{Kain:2010rb, RindlerDaller:2009er, RindlerDaller:2011kx}). Here we are especially interested in rotating BEC of short-range order and will mainly focus on non-spherical configurations that are specified by a fixed amount of angular momentum in the context of scalar dark matter clumps. To analyze these clumps we will utilize a variational approach. The variational method is often used in the study of BECs due to the complexity of equations of motion involved; see, for example, Refs.~\cite{Guth:2014hsa, Chavanis:2011zi} for spherically symmetric BECs and Ref.~\cite{Sarkar:2017aje} for rotating BECs.

Since angular momentum is a conserved quantity, the system will not be able to readily shed this quantity. An important question is: what establishes the value of angular momentum in the first place? A full answer to this may require a simulation from the early universe to determine how much angular momentum can be maintained in different patches of the universe. Of course the total angular momentum of the universe is very small, if not zero. But local patches generically carry non-zero angular momentum, as is observed for essentially all astrophysical bodies, such as stars, planets, galaxies, etc. So we find it at least plausible that some fraction of these scalar (axion) clumps would carry appreciable angular momentum.  In any case, we find that the maximum mass of the clumps in the attractive case is raised and the minimum radius in the repulsive case is also raised relative to the zero angular momentum case described above. We find solutions at both small and large angular momentum, and describe how a claim in the literature in Ref.~\cite{Davidson:2016uok} of a maximum amount of angular momentum is invalid.

This leads to a second motivation for studying these non-spherical configurations: In a forthcoming paper \cite{Hertzberg2018} we shall investigate the possibility of resonance of axion clumps from the axion-photon-photon coupling $\Delta\mathcal{L}\propto \phi\,{\bf E}\!\cdot\!{\bf B}$, as well as coupling to hidden sector photons. We shall show that whether the resonance is present or shut-off is sensitive  to the spatial size and density of the clump. Since we find that the radius is increased (at fixed mass) as we increase the angular momentum (just as the eigenstates of the hydrogen atom have larger radii) and the maximum mass is increased, it alters the resonance structure.

The outline of this paper is as follows: 
In Section \ref{AxionFieldTheory} we describe the basics of scalar field theory and take the non-relativistic limit. 
In Section \ref{NonSpherical} we introduce a class of configurations that carry angular momentum. 
In Section \ref{AttractiveSelf} we compute the properties of clumps with attractive self-interactions.
In Section \ref{RepulsiveSelf} we compute the properties of clumps with repulsive self-interactions.
In Section \ref{Astrophysical} we discuss possible astrophysical implications of our results. 
Finally, in Section \ref{Conclusions} we discuss our results.

\section{Classical Field Theory}\label{AxionFieldTheory}

\subsection{Scalar Fields and Axions}\label{Scalars}

Consider a real scalar field $\phi$ that is rather light; sufficiently light that it must be in the high occupancy regime to comprise the dark matter. It is governed accurately by classical field when an appropriate ensemble averaging is performed \cite{Hertzberg:2016tal}. Perhaps the most important example is the QCD axion. This is a pseudo-Goldstaone boson from a spontaneously broken PQ symmetry, and provides a plausible solution to the strong CP problem in the Standard Model \cite{Peccei:1977hh,Weinberg:1977ma,Wilczek:1977pj}. 

At low energies we can readily focus on the two derivative action. By operating in the Einstein frame, the action can always be put in the canonical form (units $\hbar=c=1$, signature + - - -)
\beq
\mathcal{L} = \sqrt{-g}\left[{\mathcal{R}\over 16\pi G}+\frac{1}{2} g^{\mu\nu}\nabla_{\mu}\phi\, \nabla_{\nu}\phi - V(\phi)\right]\,.
\label{axionlagrangiandensity}
\eeq
The specific form of the potential $V$ is model dependent. For the QCD axion it arises from QCD instantons which are non-trivial to compute with accuracy. The classic dilute gas approximation gives the potential
\beq
V(\phi)= {m_u m_d\over (m_u+m_d)^2}\,f_\pi^2 m_\pi^2 \left[1-\cos\left(\phi\over f_a\right)\right]\,,
\label{dilute}\eeq
where $m_{u,d}$ are the up/down quark masses, $m_\pi$ is the pion mass, $f_\pi$ is the pion decay constant, and $f_a$ is the PQ symmetry breaking scale. However, other computations have challenged the accuracy of the dilute gas approximation, leading to a more accurate estimate for the potential (e.g., see \cite{diCortona:2015ldu})
\beq
V(\phi)=f_\pi^2 m_\pi^2\left[1-\sqrt{1-{4m_um_d\over(m_u+m_d)^2}\sin^2\!\left(\phi\over2f_a\right)}\right]\,.
\label{Vimp}\eeq

We shall be interested in the behavior at small field values, where we can expand either of these potentials as
\beq
V(\phi)={1\over2}\ma^2\,\phi^2+{\lambda\over 4!}\,\phi^4+\ldots\,.
\eeq
In both of these potentials, the axion mass is given by
\beq
\ma^2={m_um_d\over(m_u+m_d)^2}{f_\pi^2 m_\pi^2\over f_a^2}\,,
\eeq
while the (negative) quartic coupling $\lambda$ differs in the two approximations. We can write it as
\beq
\lambda = -\gamma {\ma^2\over f_a^2}<0,
\label{lambdachoice}\eeq
where $\gamma=1$ in the dilute gas approximation Eq.~(\ref{dilute}) and $\gamma=1-3m_um_d/(m_u+m_d)^2\approx0.3$ in the non-dilute gas treatment Eq.~(\ref{Vimp}).

The axion is expected to be initially displaced from its potential minimum. Then at the QCD phase transition, the axion begins to oscillate back and forth in this potential, red-shifting and acting as a form of cold dark matter \cite{Preskill:1982cy,Abbott:1982af,Dine:1982ah,Kim:2008hd}. This is known as the misalignment mechanism. Although there are complications arising from the scale of inflation \cite{Fox:2004kb,Hertzberg:2008wr}, the usual bound on the symmetry breaking scale is $f_a\lesssim 10^{12}$\,GeV to avoid too much axion dark matter. As a concrete example, we will illustrate our final results with representative parameters $\ma=10^{-5}$\,eV, with $f_a = 6 \times 10^{11}\,\text{GeV}$. In this regime, the axion can comprise a significant component, or all, of the dark matter. In principle topological defects can form \cite{Sikivie:1982qv,Vilenkin:1982ks,Kim:1986ax,Barr:1986hs}, such as cosmic strings, although they are unstable. For multiple minima in the potential there can be domain walls, which are problematic in this post-inflationary scenario, so we shall assume only a single minimum of the potential here.

\subsection{Non-Relativistic Field Theory}

By now it is standard to take the non-relativistic limit of a scalar field theory, which we briefly recapitulate here. A rigorous treatment is given in Ref.~\cite{Namjoo:2017nia}, which includes the leading relativistic corrections to the non-relativistic theory. However, only the leading order, or Galilean, terms will be of significance to us here. The reason is the following: For attractive self-interactions, a maximum mass for the clumps will arise when the self-interaction energy is of the order of the gravitational energy and is also of the order of the kinetic energy. This will result in a maximum speed (we illustrate the spherically symmetric case for simplicity here; see Section \ref{Validity} for a related treatment of the non-spherical case)
\beq
v_{max}^2\sim \delta\equiv {G\,\ma^2\over|\lambda|}={G\,f_a^2\over\gamma},
\eeq
(where $G$ is Newton's gravitational constant) and even smaller speeds for more dilute clumps. Then so long as $f_a$ is sub-Planckian (such as $f_a\sim 10^{12}$\,GeV, as mentioned above, giving $\delta\sim 10^{-14}$) this characteristic speed is much smaller than the speed of light. Furthermore, on the stable branch of solutions, the field amplitudes are always small, which means the potential $V$ is dominated by the mass term. So the frequency will be close to the mass $\ma$ plus small corrections. This all fits into the non-relativistic regime.

In this non-relativistic limit, it is standard to re-express the real field $\phi$ in terms of a complex scalar field $\psi$ as  
\beq
\phi({\bf{x}},t)=\frac{1}{\sqrt{2\ma}}\left[e^{-i\ma t}\psi({\bf{x}},t)+e^{i\ma t}\psi^{*}({\bf{x}},t)\right]\,.
\label{phi}
\eeq
As the scalar field $\phi$ oscillates in its potential, its oscillation frequency will be close to $\ma$ if the amplitude of oscillations are small. This is captured by factoring for the $e^{\pm i\ma t}$ terms. Then the (important) corrections to this simple time dependence is captured in the complex field $\psi$, which is assumed to be slowly varying. For completeness, one should also demand that the momentum conjugate for $\phi$ is uniquely determined by $\psi$ and $\psi^*$ to avoid the introduction of additional degrees of freedom. However we will not elaborate on those details here. 

It suffices to insert this form for $\phi$ into the Lagrangian density Eq.~(\ref{axionlagrangiandensity}). Then any terms that are proportional to (powers of) $e^{-i\ma t}$ and $e^{i\ma t}$ can be ignored since they time average towards zero in the non-relativistic limit. Also we ignore $\sim|\dot\psi|^2/\ma$ terms relative to $\sim i\dot\psi\psi$ terms in the kinetic part of Eq.~(\ref{axionlagrangiandensity}). Finally, we write the metric in Newtonian gauge, where only the time-time component is important in the non-relativistic regime $g_{00}=1+2\,\phi_N(\psi^*,\psi)$. Then we obtain the following non-relativistic Lagrangian density for $\psi$, $\psi^*$, and $\phi_N$ 
\beq
\mathcal{L}_{nr} = \frac{i}{2}\left(\dot{\psi}\psi^{*}-\psi\dot{\psi}^{*} \right)-\frac{1}{2m}\nabla \psi^*\! \cdot\! \nabla \psi -V_{nr}(\psi,\psi^*) - m\,\psi^*\psi\,\phi_N(\psi^*,\psi)- {1\over8\pi G}(\nabla\phi_N)^2\,.
\label{nonrelativisticlagrangiandensity}
\eeq
The effective potential in this regime arises primarily from the leading quartic interaction
\beq
V_{nr}(\psi,\psi^*) = {\lambda\over16}{\psi^{*2}\psi^2\over \ma^2}\,.
\eeq
This suffices for $\phi\ll \ma/\sqrt{|\lambda|}\sim f_a$, which is a requirement of the non-relativistic approximation (we shall check on this condition in Section \ref{Validity}). 

Large field values $\phi\gtrsim \ma/\sqrt{|\lambda|}\sim f_a$ lead to significant relativistic corrections. One consequences is that there are appreciable particle number changing processes, including $4\,\phi\to2\,\phi$, etc. Such effects cause the clumps to radiate into (semi)relativistic axions, and quick evaporation (and collapse). This occurs for so-called axitons that can form in the very early universe \cite{Kolb:1993hw}. Such configurations therefore do not last long; further discussion on this subject includes Refs.~\cite{Schiappacasse:2017ham,Visinelli:2017ooc}. Hence in order to focus on only the long lived configurations, it is justified to focus on small field amplitudes, where the particle number changing processes are suppressed. Associated with this is that the above action has acquired an (accidental) global $U(1)$ symmetry $\psi\to\psi\,e^{i\theta}$. The conserved number associated with this is
\beq
N=\int d^3x\,\psi^*({\bf x})\psi({\bf x})\,,
\eeq
(technically this has units of energy$\times$time in classical field theory, and so should be multiplied by $1/\hbar$ to determine the correspond particle number in the quantum field theory).

\subsection{Hamiltonian Formulation}

In this paper we will focus on extremizing the energy (at fixed particle number, and fixed angular momentum, as we will explain). So it is important to pass to the Hamiltonian representation. In this non-relativistic limit, the momentum conjugate for $\psi$ is simply 
\beq
\pi = {\partial\mathcal{L}_{nr}\over\partial\dot\psi}=i\,\psi^*\,,
\eeq
(after performing an integration by parts). So it will suffice to represent the Hamiltonian directly in terms of $\psi$ and $\psi^*$, with the understanding that they are conjugate to each other (up to a factor of $i$). 

Note that the Newtonian potential $\phi_N$ is non-dynamical. So we can eliminate it by solving its constraint equation, which is the standard Newton-Poisson equation
\beq
\nabla^2\phi_N=4\pi G\,\ma\,\psi^*\psi\,.
\eeq
Solving this and passing to the Hamiltonian, we readily find the following 3 well known contributions to the energy \cite{Schiappacasse:2017ham,Guth:2014hsa}
\beq
H_{nr} = H_{kin} + H_{int} + H_{grav}\,,
\label{hamiltoniandensityschematic}
\eeq
where 
\bea
H_{kin} \amp=\amp {1\over2\ma}\int d^3x\, \nabla \psi^*\! \cdot\! \nabla \psi\,, \label{Hkin}\\
H_{int} \amp=\amp \int d^3x \, V_{nr}(\psi,\psi^*) \,,\label{Hint}\\
H_{grav} \amp=\amp -\frac{G\,\ma^2}{2}\int d^3x  \int d^3x' \frac{\psi^*({\bf{x}})\psi^*({\bf{x}}')\psi({\bf{x}})\psi({\bf{x}}')}{|{\bf{x}}-{\bf{x}}'|}\,.
\label{Hgrav}
\eea
These terms $H_{kin}$, $H_{int}$, $H_{grav}$, are the kinetic energy, self-interaction energy, and gravitational energy, respectively. The Hamilton equation $\dot\pi=-\delta H/(\delta\psi)$ readily leads to the following equation of motion
\beq
i\,\dot\psi=-{\nabla^2\psi\over2\ma}-G\ma^2\,\psi\!\int d^3x'\frac{\psi^*({\bf{x}}')\psi({\bf{x}}')}{|{\bf{x}}-{\bf{x}}'|}+{\partial\over\partial\psi^*}V_{nr}(\psi,\psi^*)\,,
\label{EqMotion}
\eeq
where the self-interaction term is $\partial V_{nr}/\partial\psi^* = \lambda\,\psi^*\psi^2/(8\,\ma^2)$.

\section{Including Angular Momentum}\label{NonSpherical}

In our previous work Ref.~\cite{Schiappacasse:2017ham} we analyzed the above Hamiltonian in the context of spherically symmetric clump configurations. This includes the true ground states (BEC) of the system at fixed particle number. Here we would like to generalize this work by allowing for some non-zero angular momentum. Hence we will minimize the energy at fixed particle number and fixed angular momentum. Since we are still interested in stationary solutions, one can readily show that any such solutions must have the following time and space dependence
\beq
\psi({\bf x},t) = \Phi({\bf x})\, e^{-i\,\mu\,t}\,,
\label{stationary}\eeq
where $\mu$ is the chemical potential and $\Phi({\bf x})$ may in general be a complicated function of position.

\subsection{Single Spherical Harmonic Ansatz}

A general non-spherical solution of the equations of motion is a complicated function that can be expressed as a sum over all spherical harmonics $Y_l^m(\theta,\varphi)$, with standard integer values
\beq
l=0,1,2,3,\ldots\,\,\,\,\mbox{and}\,\,\,\,\, m=-l,-l+1,\ldots,l-1,l\,.
\eeq
However, as a simple ansatz, we shall consider a field configuration specified by a single spherical harmonic 
\beq
\Phi({\bf x}) =\sqrt{4\pi}\,\Psi(r)\,Y_l^{m}(\theta,\varphi)\,,
\label{trialangular}
\eeq
where $\Psi(r)$ is some radial profile. Although non-linearities will couple different spherical harmonics to one another, we shall ignore such corrections here. In this ansatz, the total angular momentum 
\beq
L_c={1\over2}\epsilon_{abc}\int d^3x\left[i\,x^a\,\psi\,\partial^b\psi^*-i\,x^a\,\psi^*\,\partial^b\psi\right]\,,
\eeq
is readily determined to be
\beq
{\bf L}=(0,0,N m)\,,
\label{AngMom}
\eeq
with particle number $N=4\pi\int_0^\infty dr\, r^2|\Psi(r)^2|$. Note that ${\bf L}$ only depends on the spherical harmonic number $m$ and not $l$. While at first sight this may seem surprising, it is readily understood as follows: In quantum mechanics, the expectation value of the square of the angular momentum operator $\langle\psi| {\bf L}^2 |\psi\rangle $ in a state in which all $N$ particles are in the same angular momentum eigenstate is
\beq
\langle\psi| {\bf L}^2 |\psi\rangle = N\, l(l+1)+N(N-1)m^2\,.
\eeq
So for $N=1$ this recovers the familiar result $\langle\psi| {\bf L}^2 |\psi\rangle = l(l+1)$. On the other hand, as we send $N\to\infty$, this becomes $\langle\psi| {\bf L}^2 |\psi\rangle \approx N^2m^2$, which matches the classical field result of Eq.~(\ref{AngMom}). 

In this work, we will be interested in finding configurations that extremize the energy, subject to the constraint that it is at a fixed particle number (as before) and also at a fixed angular momentum. We shall see that this condition will enforce $l=|m|$.

By inserting the single spherical harmonic ansatz of Eq.~(\ref{trialangular}) into the non-relativistic Hamiltonian, we obtain the following expressions for the kinetic energy and self-interaction energy
\bea
H_{kin} \amp = \amp 4\,\pi\!\int_0^\infty dr\,r^2\left[{1\over2\,\ma}\!\left(d\Psi\over dr\right)^2+{l(l+1)\over2\,\ma \,r^2}\Psi^2\right]\,,\label{HkinAng}\\
H_{int} \amp = \amp 4\,\pi\sum_{l'=0}^{2l}(2l'+1)\,\Clm(l')\int_0^\infty dr\,r^2\,V_{nr}(\Psi,\Psi)\,,\label{HintAng}
\eea
where the coefficients $\Clm(l')$ arise from integrals over spherical harmonics and can be expressed in terms of the Wigner 3-j symbols as
\beq
\Clm(l') = (2l+1)^2\binom {l~~l'~~l} {0~~0~~0}^{\!2} \binom {~l~~l'~~l} {\!-m~0~~m}^{\!2}\,,
\eeq
which is only non-zero for {\em even} integers $l'$ in the domain $0\leq l'\leq 2l$.
In order to compute the gravitational energy it is useful to use the inverse distance expansion
\beq
\frac{1}{|{\bf x}-{\bf x}'|}= \sum_{l=0}^{\infty}\frac{4\pi}{2l+1}\left( \frac{r_<^l}{r_{>}^{l+1}} \right) \sum_{m=-l}^{l}Y_{l}^{{m}^*}(\theta,\varphi)Y_{l}^m(\theta',\varphi')\,,
\label{inversedistanceexpansion}
\eeq
where $r_<$ is the lesser and $r_>$ is the greater of $r=|{\bf x}|$ and $r'=|{\bf x}'|$. This leads to the following expression for the gravitational energy
\bea
H_{grav} \amp = \amp -{G\,\ma^2\over2}(4\pi)^2\sum_{l'=0}^{2l}\Clm(l')\!\int_0^\infty dr\,r^2\!\int_0^\infty dr'\,r'^2{\Psi(r)^2\Psi(r')^2\over r_{>}}\left(r_{<}\over r_{>}\right)^{\!l'}\,,\label{HgravAng}
\eea
where again the coefficients $\Clm(l')$ appear.

\subsection{Modified Gaussian Ansatz}

To make analytical progress it is useful to have an ansatz for the radial profile $\Psi(r)$. In our previous work \cite{Schiappacasse:2017ham} we considered various ansatzes, including an exponential ansatz $\Psi\propto \exp(-r/R)$, a sech ansatz $\Psi(r)\propto \mbox{sech}(r/R)$, and a modified exponential ansatz $\Psi(r)\propto (1+r/R)\exp(-r/R)$. The latter ansatzes were found to be especially accurate when compared to the exact numerical results. In fact in the latter two ansatzes, the small field expansion $\Psi(r)\propto 1-r^2/(2R^2)+\ldots$ was necessary to be consistent with the equation of motion. However, for non-zero $l$, the structure of the small $r$ behavior is drastically altered compared to the $l=0$ case. The time independent scalar field equation at small $r$ is
\beq
\mu_{eff}\,\Psi\approx -{1\over 2\ma}\left(\Psi''+{2\over r}\Psi'\right)+{l(l+1)\over2\ma\,r^2}\Psi\,\,\,\,\,(\mbox{near region})\,,
\label{nearregionschrodingerAng}
\eeq
where we allow for a possible shift in the chemical potential $\mu\to\mu_{eff}$ to account for the gravitational term. As $r\to0$ we have the potential problem that this equation will blow up due to the $\sim l(l+1)\Psi/r^2$ and $\sim \Psi'/r$ terms. To avoid this problem, we need to find a $\Psi$ such that the $\sim l(l+1)\Psi/r^2$ and $\sim \Psi'/r$ divergences cancel in this limit. It is straightforward to check that this requires
\beq
\Psi(r)=\Psi_{\alpha}\,r^l-{1\over 2}\Psi_{\beta}\,r^{l+2}+\ldots\,\,\,\,\,(\mbox{near region})\,,
\label{SmallrAng}
\eeq
where $\Psi_{\alpha}$ and $\Psi_{\beta}$ are constants. 

The medium to large $r$ regime is rather complicated and requires an ansatz. Numerical experimentation has revealed that a surprisingly accurate ansatz for non-zero $l$ is a modified Gaussian profile
\beq
\Psi_R(r)=\sqrt{N\over2\pi(l+{1\over2})! \, R^3}\,\left(r\over R\right)^{\!l}\,e^{-r^2/(2 R^2)}\,\,\,\,\,(\mbox{modified Gaussian ansatz})\,,
\label{ModGauss}
\eeq
which
\begin{figure}[t]
\centering
\includegraphics[scale=0.35]{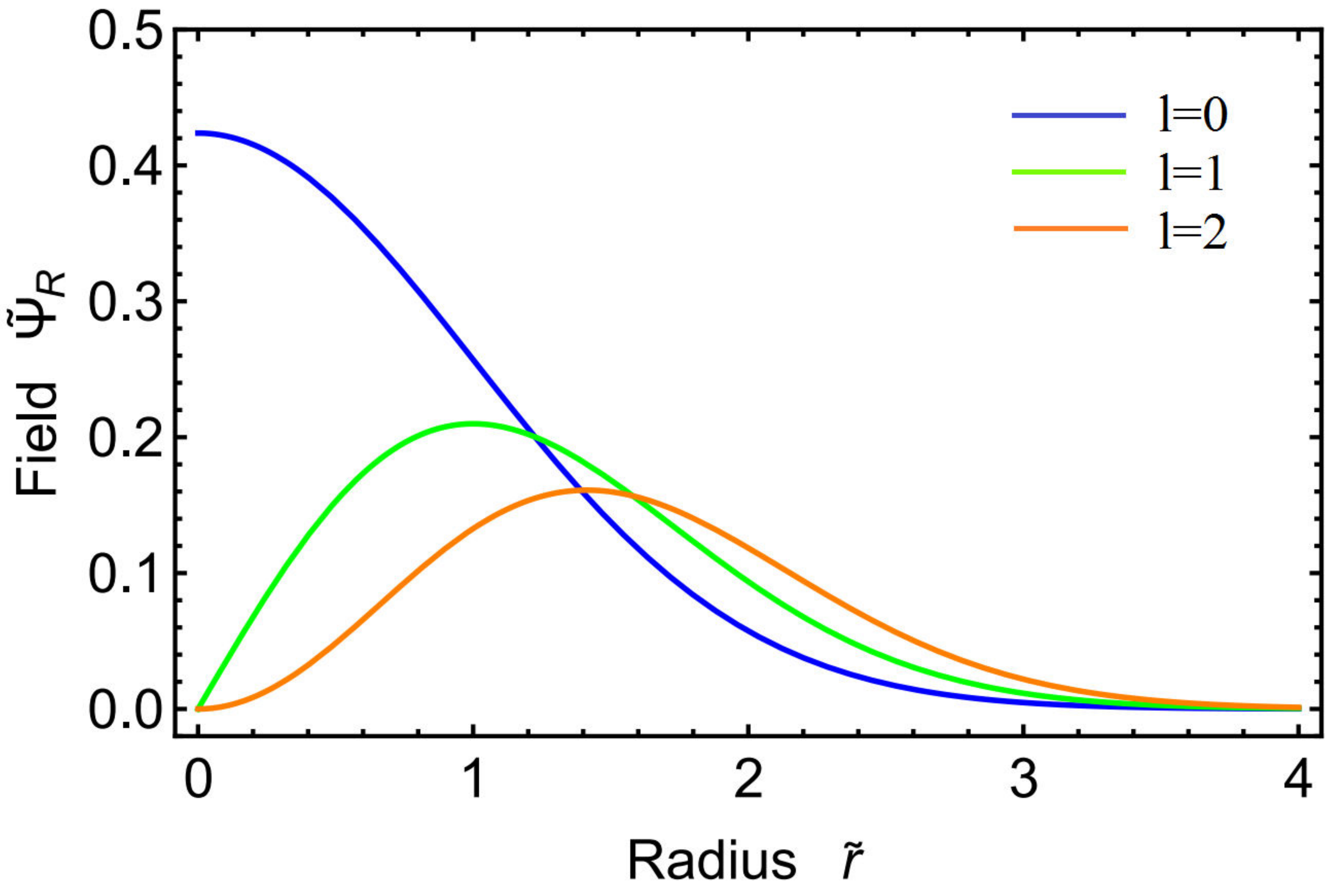}
\caption{Field $\tilde{\Psi}_R=\Psi_R\sqrt{R^3/N}$ versus radius $\tilde{r}=r/R$ in the modified Gaussian ansatz for different values of spherical harmonic number $l$. Blue is $l=0$, green is $l=1$, and orange is $l=2$.}
\label{FieldvsRadius}
\end{figure}
evidently has the desired properties of Eq.~(\ref{SmallrAng}) for small $r$. Here the prefactor is specified uniquely in order to ensure correct normalization $N=4\pi\int dr\,r^2\,\Psi(r)^2$. The first few, namely $l=0, 1,2$, are plotted in Fig.~\ref{FieldvsRadius}.
Here the radius $R$ is a variational parameter, and will be adjusted to minimize the energy accordingly. In general $R$ will be a function of $\{l,m\}$. 

At very large radii, the form of the equation changes to the following 
\beq
\mu\,\Psi \approx -{1\over 2\ma}\left(\Psi''+{2\over r}\Psi'\right)-{G\ma^2N\over r}\Psi\,\,\,\,\,(\mbox{far region})\,,
\label{farregionschrodinger}\eeq
which has the same structure as the Schr\"odinger equation for an atom. As is well known, the solutions for this fall off exponentially at large distances. On the other hand, our modified Gaussian ansatz Eq.~(\ref{ModGauss}) evidently falls off faster than an exponential. Nevertheless, the modified Gaussian is found to be accurate from small to moderately large $r$, which is the regime that comprises the bulk of the density. So it adequately captures the essential features of the solutions with reasonable accuracy.

\subsection{Energy vs Radius}

\begin{figure}[t]
\centering
\includegraphics[scale=0.7]{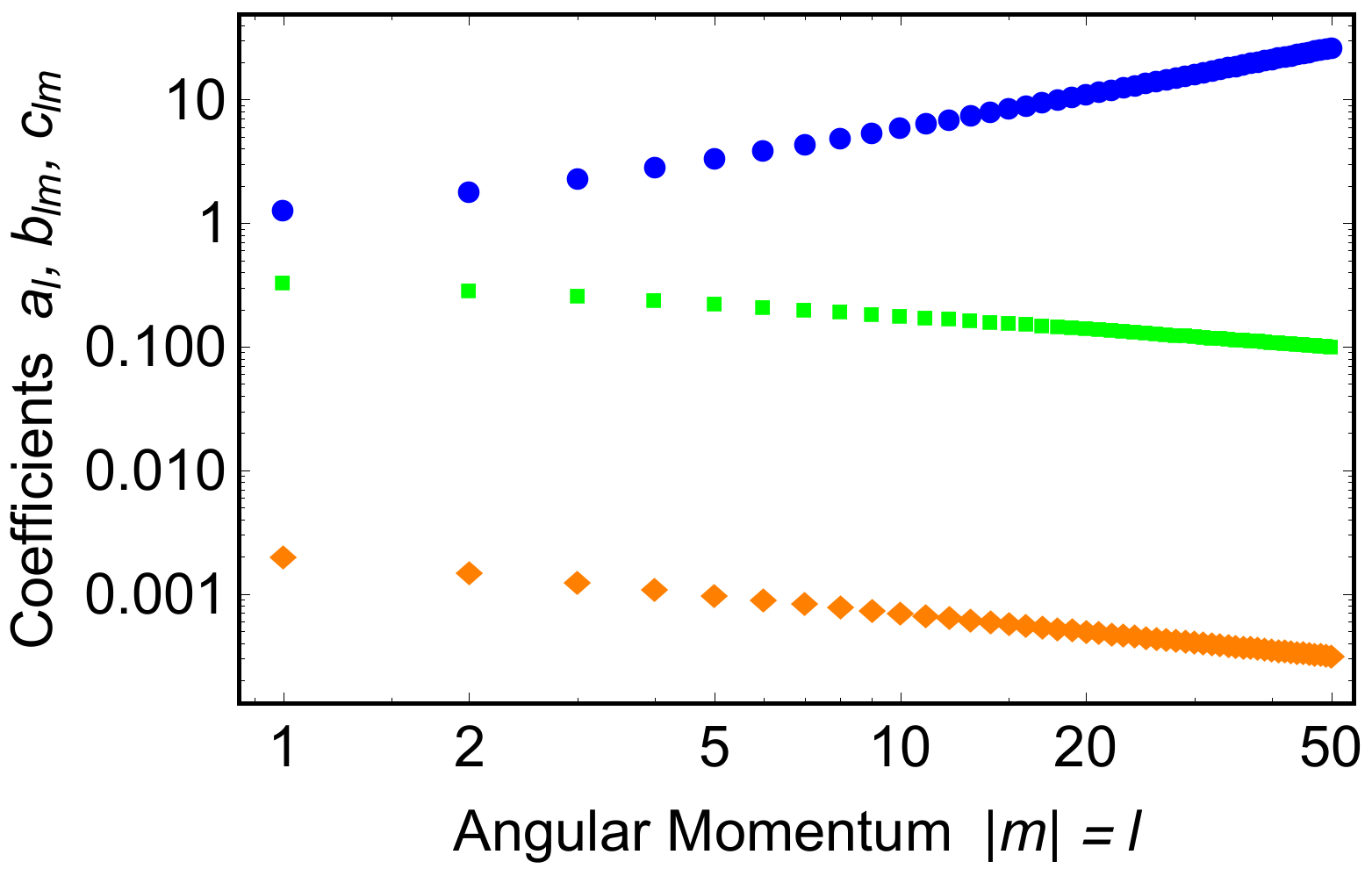}
\caption{Coefficients $a_l$ in blue (top data), $b_{lm}$ in green (middle data), and $c_{lm}$ in orange (bottom data), in the modified Gaussian ansatz, versus angular momentum parameter $|m|$, with $l=|m|$.}
\label{CoeffsAng}
\end{figure}

We insert this modified Gaussian ansatz into the full non-relativistic Hamiltonian Eqs.~(\ref{HkinAng},\,\ref{HintAng},\,\ref{HgravAng}) to obtain the energy $H$ as a function radius $R$, number $N$, and spherical harmonic numbers $\{l,m\}$. In this non-relativistic regime, we can, without loss of generality, scale out the parameters $\ma,\,G,\,|\lambda|$, by using a dimensionless radius, number, and energy, as follows
\bea
\R \amp \equiv \amp {\ma^2\sqrt{G}\over\sqrt{|\lambda|}}\,\rr \,\,\,\,\,\,\,\,\,\,\,\,\,\, \mbox{(re-scaled clump size)}\,,\label{rescR}\\
\n \amp \equiv \amp \ma\sqrt{G\,|\lambda|}\,N \,\,\,\,\,\,\mbox{(re-scaled particle number)}\,,\label{rescN}\\
\E \amp \equiv \amp {|\lambda|^{3/2}\over \ma^2\sqrt{G}}\,H_{nr}\,\,\,\,\,\,\,\,\, \mbox{(re-scaled energy)}\label{rescH}\,.
\eea
In terms of these dimensionless variables, the Hamiltonian becomes
\beq
\E(\R) = a_{l}\frac{\n}{\R^2} - b_{lm}\frac{\n^2}{\R} + c_{lm}{\lambda\over|\lambda|}\frac{\n^2}{\R^3 }\,,
\label{hamiltonianquadraticgeneralAng}
\eeq
where only the sign of $\lambda$ is important, as indicated in the final term $\sim\lambda/|\lambda|$. The coefficients $a_l,\,b_{lm},\,c_{lm}$ are positive constants, with values
\bea
a_{l}\amp = \amp {3+2l\over4}\,, \\
b_{lm}\amp = \amp \sum_{l'=0}^{2l}\Clm(l')\,J_l(l')\,, \\
c_{lm}\amp = \amp {(2l+{1\over2})!\over 2^{2l+{13\over2}}\pi\,[(l+{1\over2})!]^2}\sum_{l'=0}^{2l}(2l'+1)\Clm(l')\,.
\eea
In Fig.~\ref{CoeffsAng} we set $l=|m|$ and plot the coefficients $a_l,\,b_{lm},\,c_{lm}$ versus angular momentum number $|m|$ for $1\leq |m| \leq 50$. Note that the sign of $m$ is irrelevant here, so we indicate its absolute value.

The coefficients $J_l(l')$ arise from integrating products of the modified Gaussian ansatz, weighted by powers of $r_{>}$ and $r_{<}$. The answer is highly non-trivial, but can be expressed in terms of the hypergeometric function $_2F_1$ and the regularized hypergeometric function $_2F_1^{reg}$ as
\bea
J_l(l') \amp = \amp {(2l+{3\over2})!\, _2F_1(2l+{5\over2},l+{l'+3\over2},l+{l'+5\over2},-1)\over(3+2l+l')[(l+{1\over2})!]^2}\nonumber\\
\amp+\amp{(l-{l'\over2})!\,[(l+{l'+1\over2})!-(2l+{3\over2})!\, _2F_1^{reg}(2l+{5\over2},l-{l'\over2}+1,l-{l'\over2}+2,-1)]\over2[(l+{1\over2})!]^2}\,.
\eea

For $l=m=0$ these $a_0,\,b_{00},\,c_{00}$ coefficients describe the spherically symmetric theory within the Gaussian ansatz with $a_0=3/4,\,b_{00}=1/\sqrt{2\pi},\,c_{00}=1/(32\sqrt{2\pi^3})$. However, the Gaussian ansatz is not too accurate for $l=m=0$. This special case is better analyzed with other ansatzes, as we did previously in Ref.~\cite{Schiappacasse:2017ham}. Instead the advantage of the modified Gaussian is that it becomes more accurate as we move to aspherical configurations with non-zero values of $\{l,m\}$. 

\begin{figure}[t]
\centering
\includegraphics[scale=0.7]{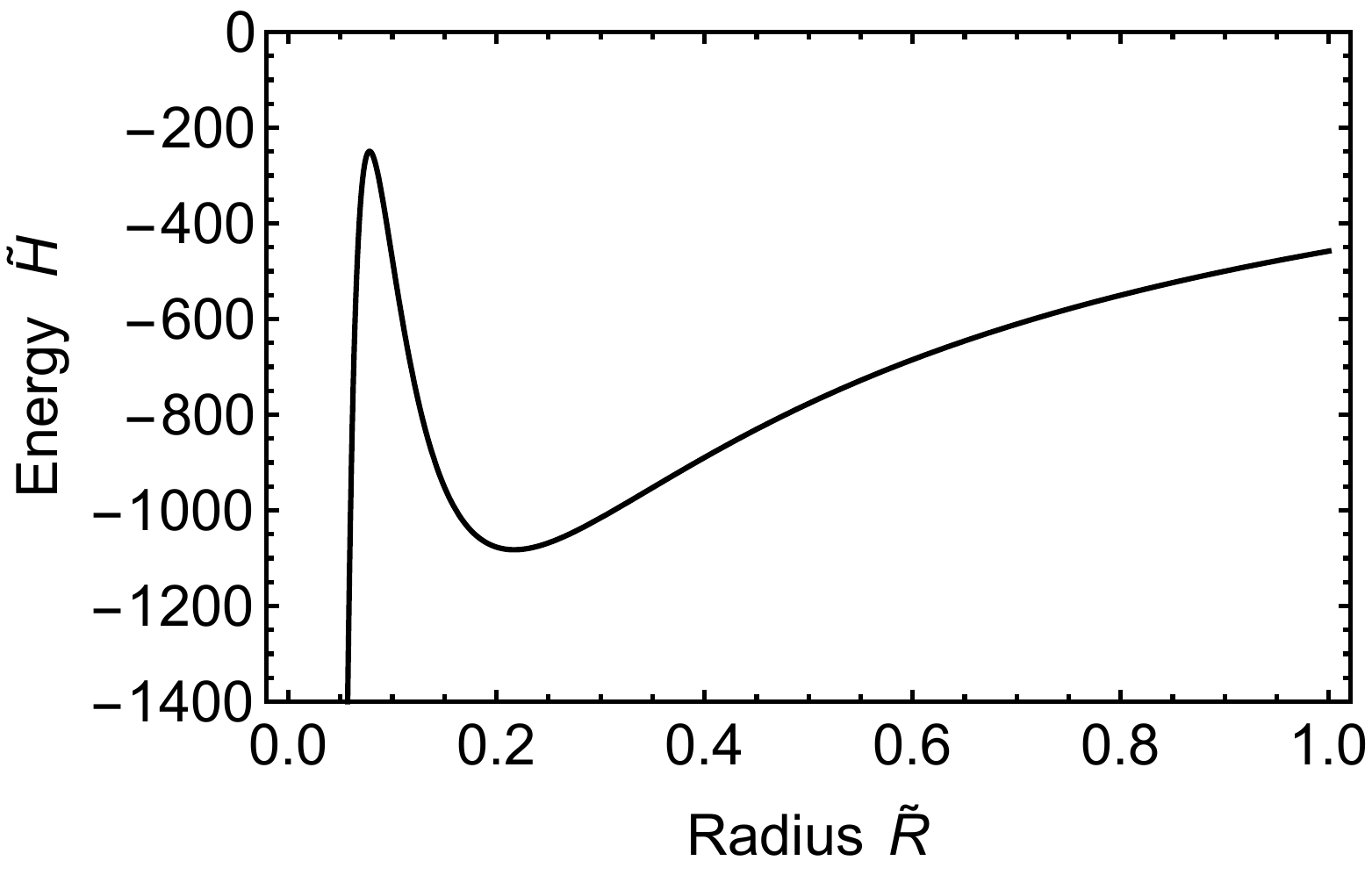}
\caption{A plot of the dimensionless energy $\E$ versus variational parameter (which is on the order of the clump radius) $\R$ for a fixed value of particle number $\n=45$ and attractive self-interactions $\lambda<0$. This is made within the modified Gaussian ansatz, with spherical harmonic numbers $l=|m|=2$. The local maximum is associated with an unstable solution and corresponds to a point on the red curve of Fig.~\ref{FigureRadiusNumberAngular}, while the local minimum is associated with a stable solution and corresponds to a point on the blue curve of Fig.~\ref{FigureRadiusNumberAngular}.}
\label{hamiltonianvsradius}
\end{figure}

\section{Attractive Self-Interactions}\label{AttractiveSelf}

\subsection{Stable and Unstable Solutions}

The energy as a function of radius is controlled in an important way by the sign of $\lambda$. Let us focus on the case of the axion for now, which is ordinarily associated with attraction $\lambda<0$ (see Eq.~(\ref{lambdachoice})). (We shall consider the repulsive $\lambda>0$ case in Section \ref{RepulsiveSelf}.) For any value of $\{l,m\}$, as long as $N$ is not too large, we find that energy as a function of radius has both a local maximum and a local minimum. This is shown in Fig.~\ref{hamiltonianvsradius}, where we plot $\E=\E(\R)$ with $\n$ fixed at $\n=45$ and spherical harmonic numbers $l=|m|=2$.  For sufficiently large $N$, there are no extrema (at fixed $\{l,m\}$).

Stationary configurations are associated with these extrema of the energy function $\partial\E/\partial\n=0$. For $\lambda<0$, this leads to the following pair of extrema
\beq
\R = {a_l\pm\sqrt{a_l^2-3\,b_{lm}\,c_{lm}\,\n^2}\over b_{lm}\,\n}\,,
\label{RabcAngular}
\eeq
with an upper bound on the number of particles
\beq
\n<\n_{max}={a_l\over\sqrt{3\,b_{lm}\,c_{lm}}}\,.
\label{NabcAngular}
\eeq
The corresponding radius, which is the minimum radius allowed for the plus-sign solution, is
\beq
\R_{min}=\sqrt{3\,c_{lm}\over b_{lm}}\,.
\label{Rmin}\eeq
The effective size of the clump is actually related to $\R$ in a non-trivial fashion. As we increase $l$ the true radius of the clump is parametrically different than $\R$. We shall define a physical radius $\rr_{90}$ by that which encloses 90\% of the configuration's mass as follows
\beq
0.9\,N=4\pi\int_0^{\rr_{90}}dr'\,r'^2\,\Psi(r')^2\,.
\eeq
Within this modified Gaussian ansatz, we have determined it numerically for each value of $l$. We note that an excellent fitting function for both small and large $l$ is found to be
\beq
\rr_{90}\approx \left(0.65+\sqrt{l+1.25}\right)\rr\,.
\label{R90}\eeq

\begin{figure}[t]
\centering
\includegraphics[scale=0.24]{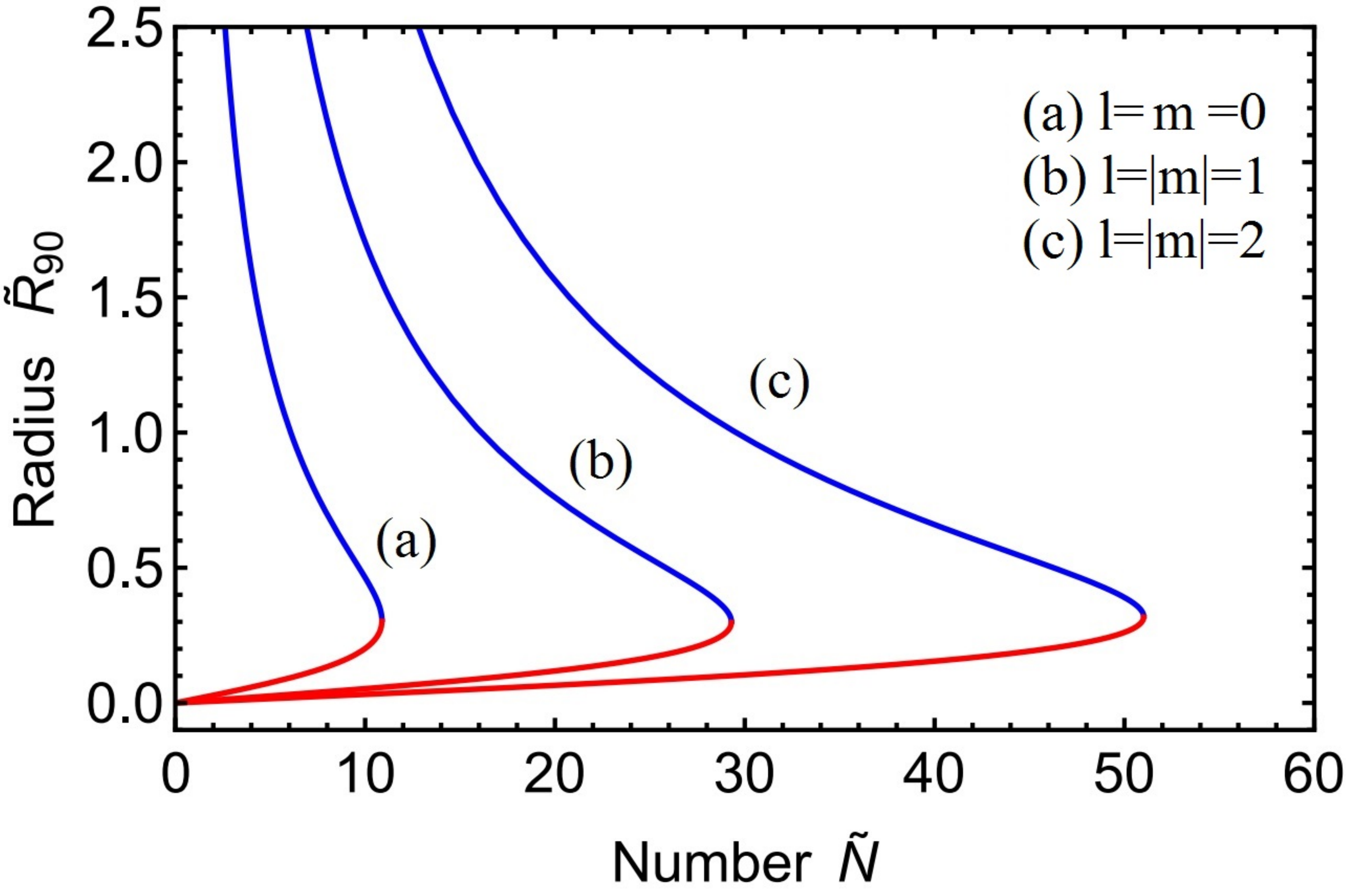}
\includegraphics[scale=0.24]{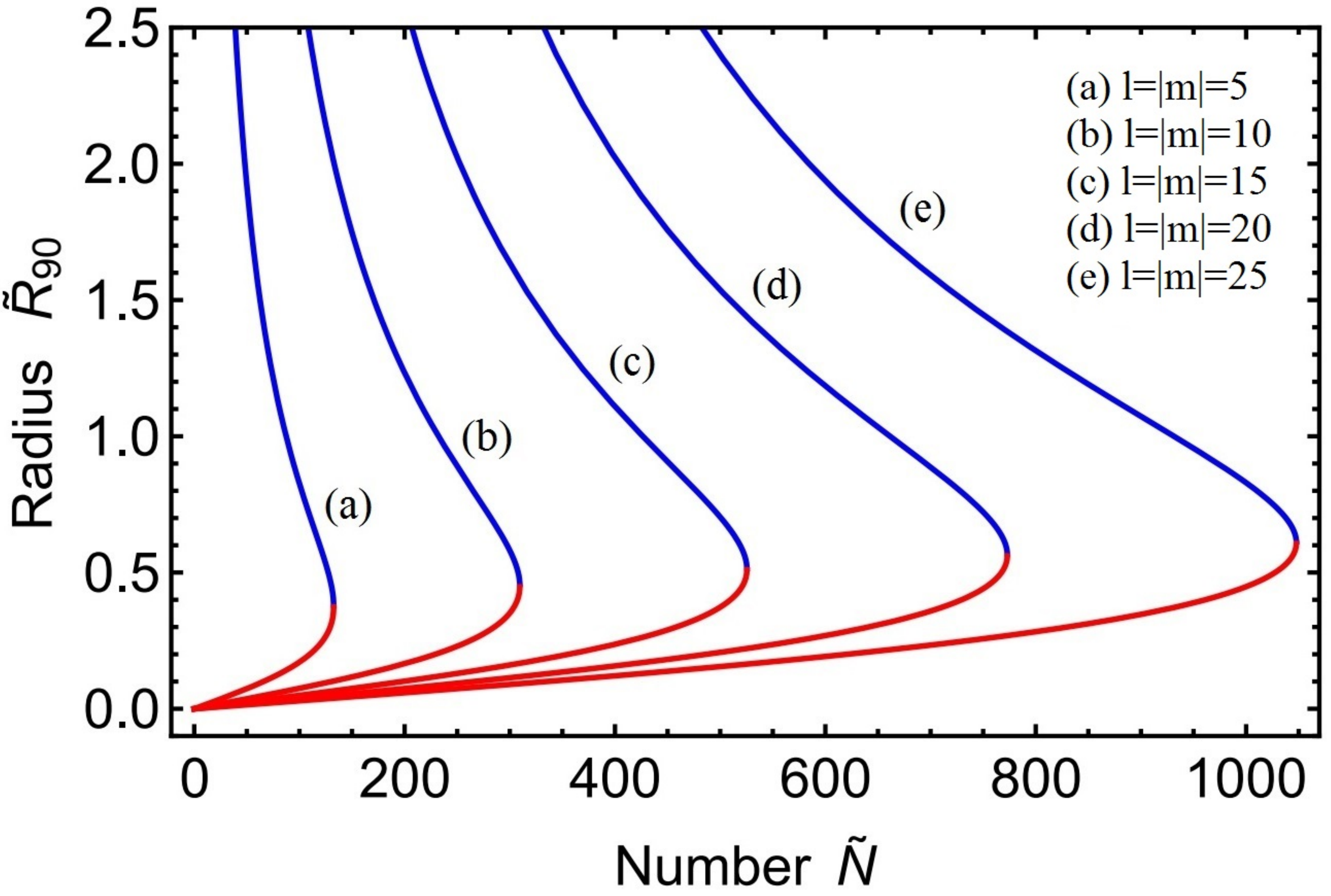}
\caption{Radius versus number for attractive (axion) clump solutions Eq.~(\ref{RabcAngular}) for different values of angular momentum within the modified Gaussian ansatz. In the left hand figure:  (a) $l=m=0$, (b) $l=|m|=1$, and (c) $l=|m|=2$. In the right hand figure: (a) $l=|m|=5$, (b) $l=|m|=10$, (c) $l=|m|=15$, (d) $l=m=20$, and (e) $l=|m|=25$. The upper blue branch corresponds to stable clump solutions, while the lower red branch corresponds to unstable clump solutions.}
\label{FigureRadiusNumberAngular}
\end{figure}

The corresponding branches of solutions are plotted in Fig.~\ref{FigureRadiusNumberAngular} for different values of $l=|m|$, using this re-scaled radius. The upper blue curves correspond to local minima of the energy function and so are associated with stable solutions. The lower red curves correspond to local minima of the energy function and so are associated with unstable solutions.

\subsection{Special Case $l=|m|$}\label{specialcase}

\begin{figure}[t]
\centering
\includegraphics[scale=0.35]{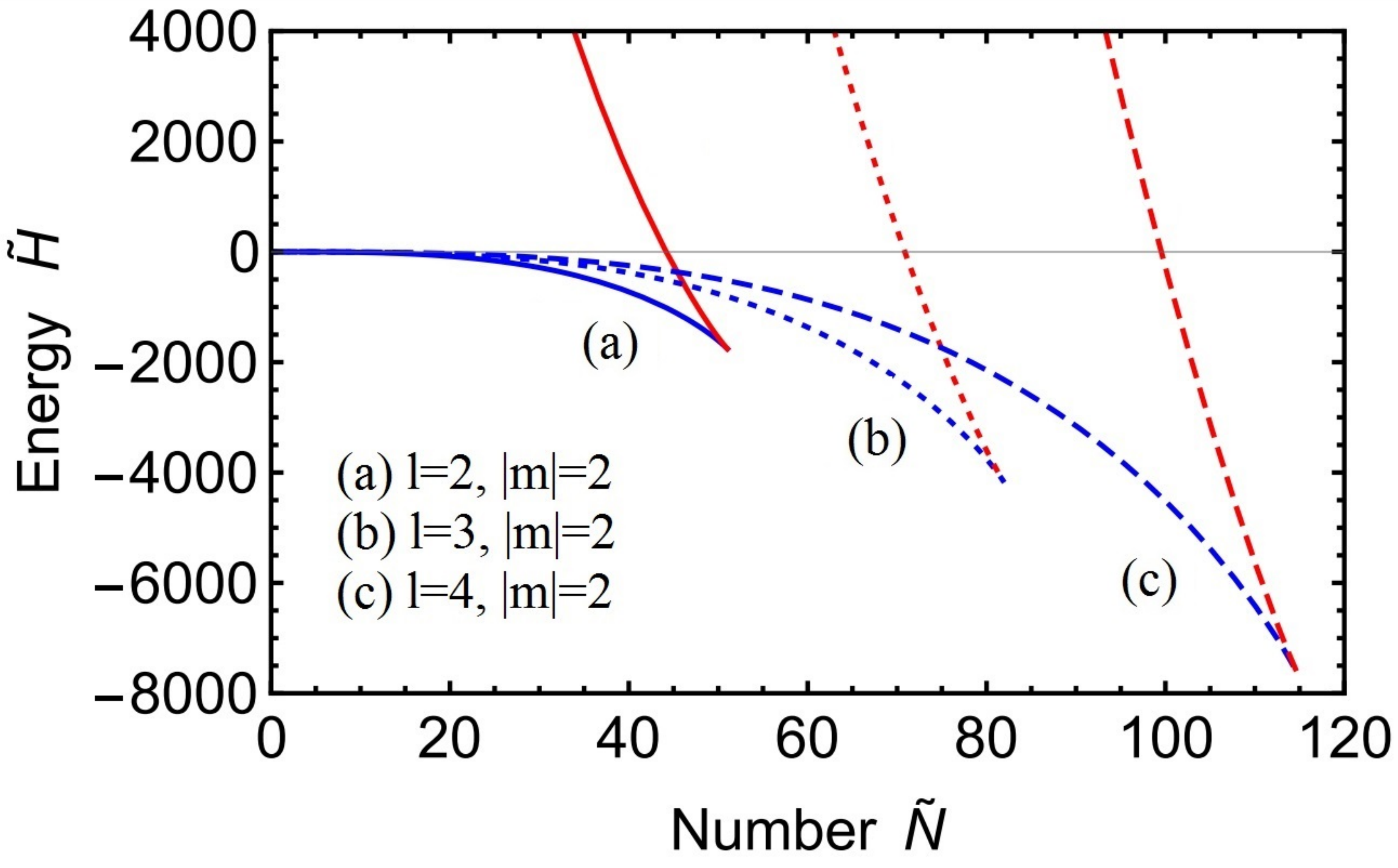}
\caption{Energy of clump solution versus number with non-zero angular momentum parameter $|m|=2$, for different values of $l$: (a) solid (lower) curve is $l=2$, (b) dotted (middle) curve is $l=3$, and (c) dashed (upper) curve is $l=4$. At a fixed number $N$ and angular momentum $L_z=N\,m$, this illustrates that the configuration that minimizes the energy has spherical harmonic number $l=|m|$ (whenever the solution exists).}
\label{FigureEnergyNumberAngular}
\end{figure}

In this work, we only need to focus on the special case
\beq
l=|m|,
\eeq 
(results are the same for $m\to-m$). The reason is the following: We are interested in a form of BEC that minimizes the energy at fixed particle number $N$ and angular momentum $L_z=N\,m$. In Fig.~\ref{FigureEnergyNumberAngular} we plot the energy $H$ on the solution branches as a function of number $N$ at a fixed amount of angular momentum number $|m|=2$ for different choices of spherical harmonic number $l=2,\,3,\,4$. We see that at a fixed $N$ (so long as the solution exists) the energy on the stable blue branch is minimized for $l=2$. In general, we find the condition for minimum energy at fixed angular momentum is $l=|m|$.

\subsection{Large Angular Momentum Limit}

Since the behavior of solutions is determined by the 3 coefficients $a_l$, $b_{lm}$, $c_{lm}$, it is important to analyze their properties, especially in the limit of large angular momentum. The expression for $a=(3+2l)/4$ is already simple, while $b_{lm}$ and $c_{lm}$ are complicated in general. Focussing on $l=|m|$, as explained above, we find that the high $l$ behavior is
\bea
b_{lm} \amp \approx \amp 0.336\sqrt{\ln l\over l} \,\,\,\,\,(\mbox{high}\,\,l=|m|)\,,\\
c_{lm} \amp \approx \amp{1\over 32\pi^2\sqrt{2\,l}}\,\,\,\,\,\,\,\,\,(\mbox{high}\,\,l=|m|)\,,
\eea
where the above expression for $c_{lm}$ is asymptotically exact in the $l\to\infty$ limit, while the above expression for $b_{lm}$ is accurate for moderate to large $l$, though not asymptotically exact in the $l\to\infty$ limit. Using Eq.~(\ref{NabcAngular}) and the exact expression for $a_l$, this leads to the following approximation for the maximum number of particles in a clump for high angular momentum $l=|m|$
\beq
\n_{max}\approx {10.52\over(\ln l)^{1\over4}}\left(l^{3\over2}+{3\over2}\sqrt{l}\right) \,\,\,\,\,(\mbox{high}\,\,l=|m|) \,.
\label{nexpand}\eeq
For example, setting $l=|m|=25$, this high $l$ estimate gives $\n_{max}\approx 1041$, to be compared to the exact numerical one given by $\n_{max}\approx 1048$. One may drop the final term in Eq.~(\ref{nexpand}) in the large $l$ limit, but for moderate $l$, it is useful to keep this term for improved precision. Similarly, the corresponding radius, which is the minimum radius allowed to remain on the stable branch, is
\beq
\R_{90,min}\approx{0.141\over(\ln l)^{1/4}}\left(0.65+\sqrt{l+1.25}\right) \,\,\,\,\,(\mbox{high}\,\,l=|m|) \,.
\label{R90high}\eeq
where we have used Eqs.~(\ref{Rmin},\,\ref{R90}).

\subsection{Comparison to Claims in Literature}

We would like to contrast these results to the work of Ref.~\cite{Davidson:2016uok}, where the authors also study rotating BECs of axion dark matter. Using the virial theorem (see the time-independent virial theorem derivation in Ref.~\cite{Chavanis:2011zi}) and an approximation scheme for the axion field, they report that the maximum mass for stable clump solutions increases with angular momentum as $\n_{max} \propto[1 + 4 l (l + 1)]/(\sqrt{l+1})$. The leading behavior at large $l$ according to this result is $\propto l^{3/2}$, which roughly agrees with our finding in Eq.~(\ref{nexpand}), up to a logarithmic correction. 

Importantly, these authors further claim there is an upper bound on the angular momentum for clump solutions to exist of $l \leq 3$. Similar to us, these authors take the profile to be a single spherical harmonic times a radial profile. They make a further simplification by only considering the angular averaged Newtonian potential; this allows them to reduce the problem to solving a pair of ODEs for the radial profile. We have taken their choice of equations and numerically solved them by searching for a radial profile that obeys the correct boundary conditions (with zero nodes). We find that as $l$ is increased, it does become numerically more intensive to find solutions. However, we have definitively found solutions for $l>3$ (we have explicitly checked various cases, such as $l=4,\,5,\,6$). In fact with sufficient precision we believe solutions exist for any $l$, as is predicted by our variational method. Furthermore, our results make sense physically: a rotating BEC that is supported by gravity with an attractive $1/r$ potential is qualitatively similar to eigenstates of the hydrogen atom with angular momentum, where we know there is no upper bound. In summary, we find no evidence to support the claim of Ref.~\cite{Davidson:2016uok} that there is such an upper bound on $l$.

\subsection{Regime of Validity}\label{Validity}

It is important to identify the regime in which we can self-consistently apply the non-relativistic approximation. This requires the field amplitude to satisfy 
\beq
\phi_0\ll{\ma\over\sqrt{|\lambda|}}\,,
\label{requirement}\eeq
in order to assure that the frequency of oscillation is close to $\ma$. Since the field has angular dependence given by $Y_l^m(\theta, \phi)$ (recall Eq.~(\ref{trialangular})), the regime of validity can depend on the choice of angular momentum. 

For stationary configurations within the single spherical harmonic ansatz (see Eqs.~(\ref{stationary},\,\ref{trialangular})) the field amplitude is given by
\beq
\phi_0 = \sqrt{2\over\ma}\,\Psi_0\,\sqrt{4\pi}\,|Y_l^l|_{max}
\eeq
where $|Y_l^l|_{max}$ is the maximum value of the spherical harmonic, which is
\beq
\sqrt{4\pi}\,|Y_l^l|_{max}\equiv g_l = {\sqrt{(2l+1)!}\over 2^l\,l!}\,.
\eeq
Also the amplitude $\Psi_0$ (maximum value) of the modified Gaussian occurs at $r=\sqrt{l}\,R$, which corresponds to 
\beq
\Psi_0 = \sqrt{N\over R^3}\,f_l,\,\,\,\,\,\mbox{with}\,\,\,\,
f_l = \sqrt{l^{l}\,e^{-l}\over2\pi(l+{1\over2})!}\,.
\eeq
The requirement in Eq.~(\ref{requirement}) to remain within the non-relativistic regime can then be translated into a lower bound on the radius $\R$ for a given number of particles $\n$ and angular momentum $|m|=l$ as
\beq
\R\gg\Rm(\n) = \left( 2\,\delta\,g_l^2 f_l^2\,\n \right)^{1/3}\,.
\eeq
Here the function $f_l$ decreases as $\sim l^{-1/2}$ at large $l$ and $g_l$ is very slowly increasing. So recalling that $\delta\equiv G\,\ma^2/|\lambda|=G\,f_a^2/\gamma\lesssim 10^{-14}$ for QCD axions in the standard window $f_a\lesssim 10^{12}$\,GeV, this criteria is readily satisfied for the entire stable blue branch of solutions, for both small and large angular momentum. It only breaks down in the lower left hand corner of the unstable red branch in Fig.~\ref{FigureRadiusNumberAngular}, which is connected to axitons, as we discussed in Ref.~\cite{Schiappacasse:2017ham}, and will not be pursued here.

\section{Repulsive Self-Interactions}\label{RepulsiveSelf}

For completeness, let us now consider the case in which the scalar self-interactions $\lambda\phi^4$ are repulsive with $\lambda>0$. This is not expected for the axion and a typical axion-like-particle, but it would occur in a standard renormalizable theory. So long as the particle remains sufficiently light, with corresponding high occupancy number to make up most or all of the dark matter, then it is amenable to the classical field theory analysis undertaken here.

\begin{figure}[t]
\centering
\includegraphics[scale=0.7]{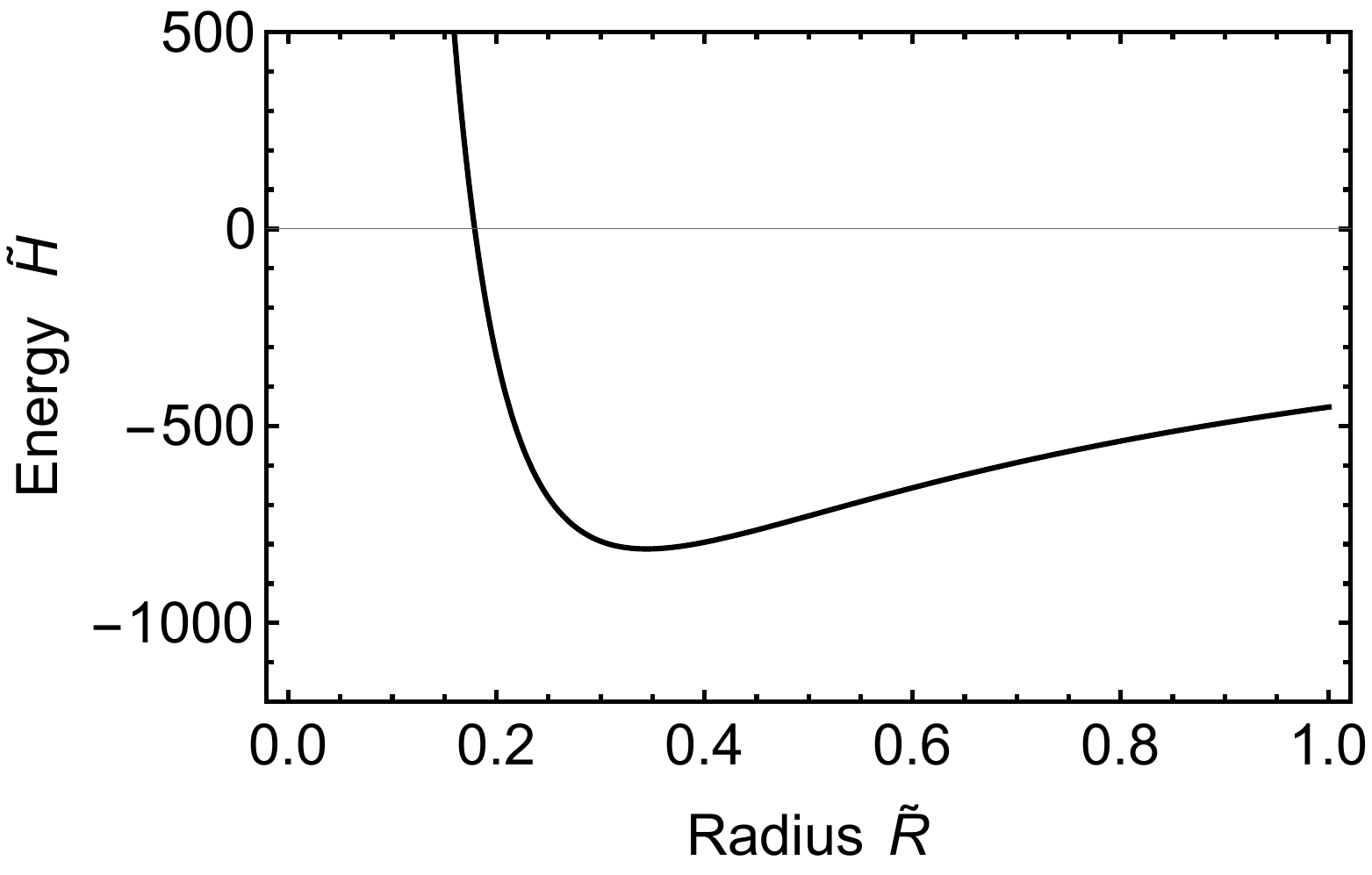}
\caption{A plot of the dimensionless energy $\E$ versus variational parameter (which is on the order of the clump radius) $\R$ for a fixed value of particle number $\n=45$ and repulsive self-interactions $\lambda>0$. This is made within the modified Gaussian ansatz, with spherical harmonic numbers $l=|m|=2$. The (global) minimum is associated with a stable solution and corresponds to a point on the blue curve of Fig.~\ref{FigureRadiusNumberRepulsive}.}
\label{hamiltonianvsradiusrepel}
\end{figure}

\begin{figure}[t]
\centering
\includegraphics[scale=0.241]{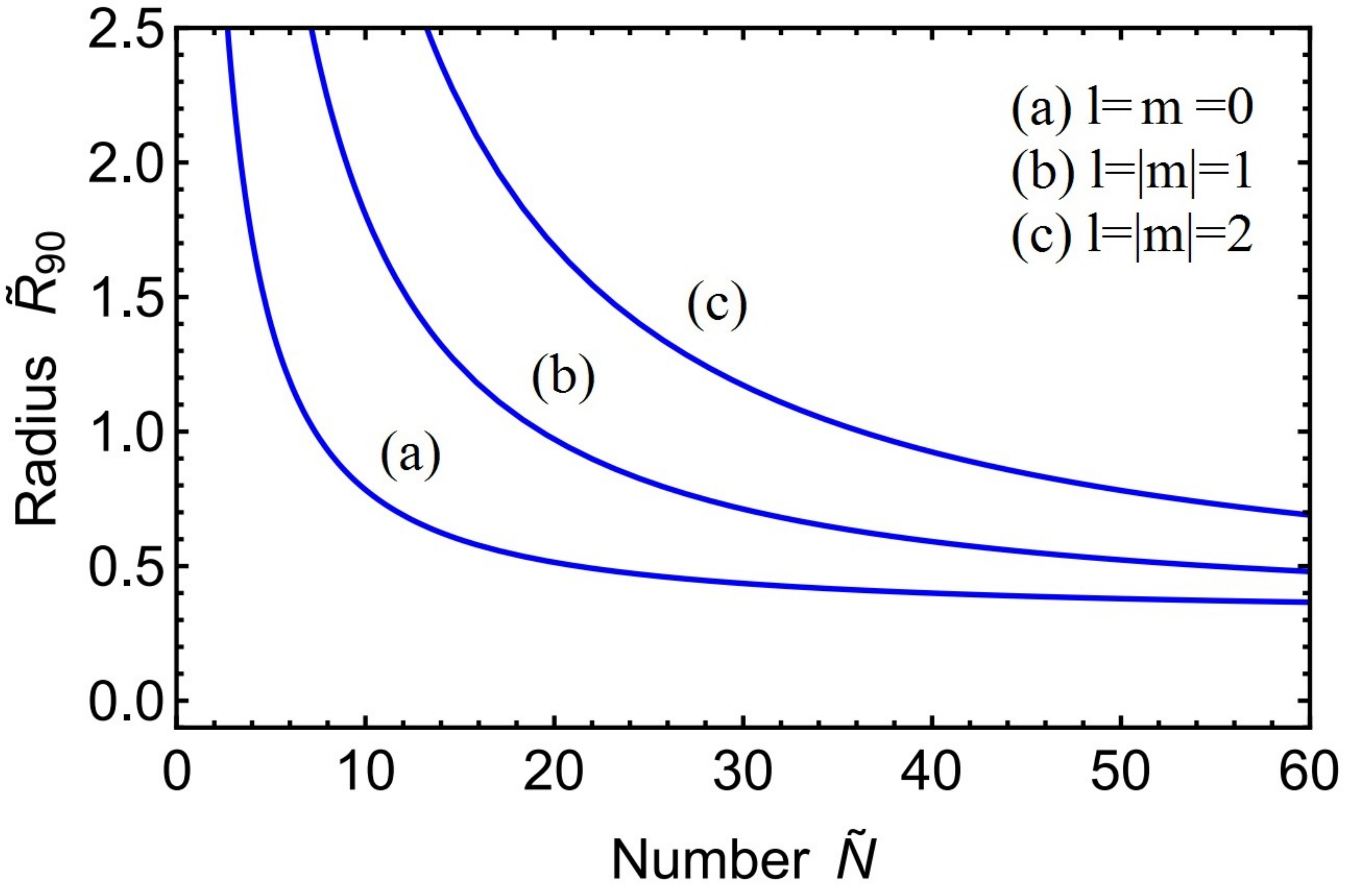}
\includegraphics[scale=0.23]{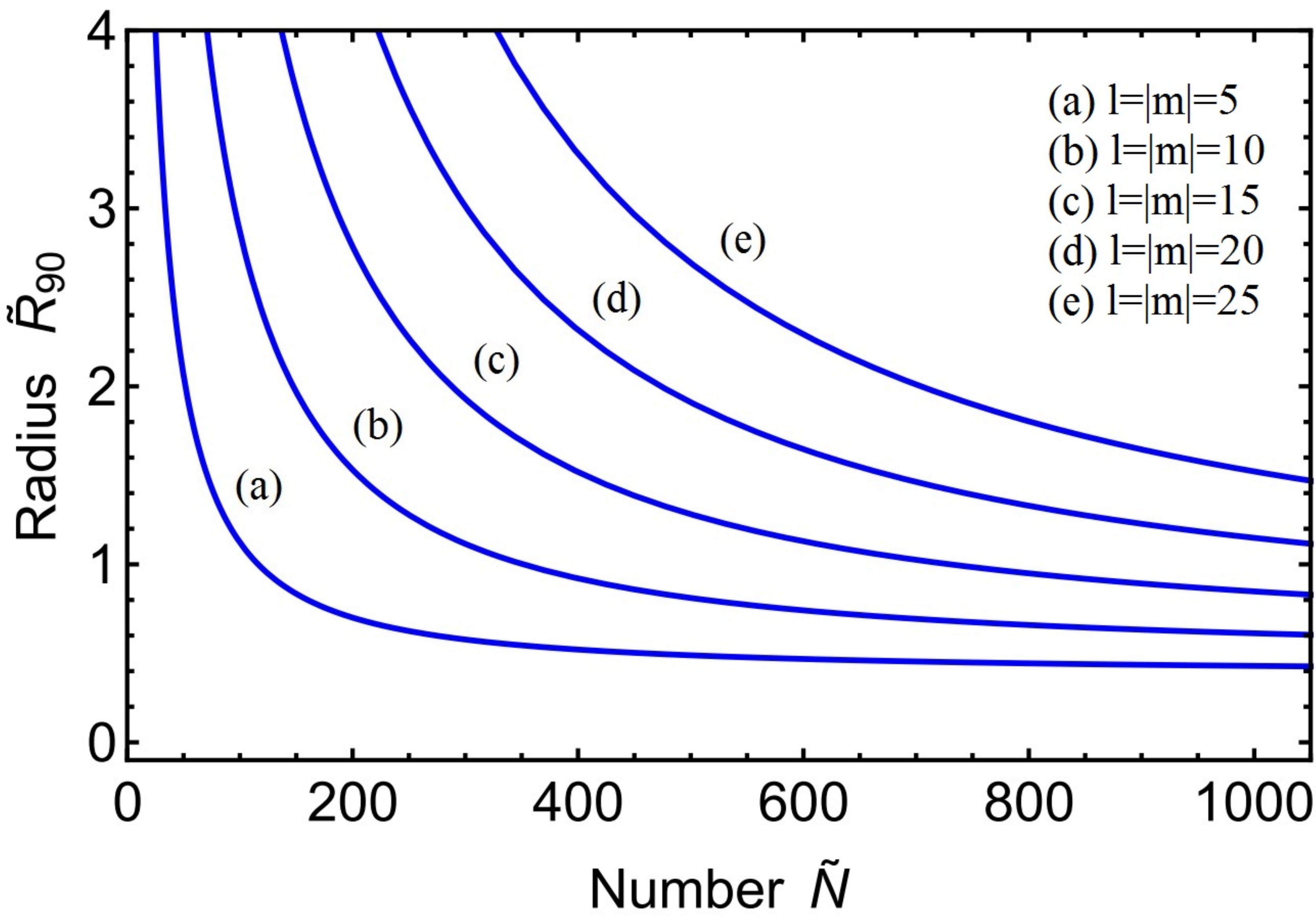}
\caption{Radius versus number for repulsive clump solutions Eq.~(\ref{Rrepel}) for different values of angular momentum within the modified Gaussian ansatz. In the left hand figure:  (a) $l=m=0$, (b) $l=|m|=1$, and (c) $l=|m|=2$. In the right hand figure: (a) $l=|m|=5$, (b) $l=|m|=10$, (c) $l=|m|=15$, (d) $l=m=20$, and (e) $l=|m|=25$. The (single) blue branch corresponds to stable clump solutions.}
\label{FigureRadiusNumberRepulsive}
\end{figure}

In the modified Gaussian ansatz, the energy formula is once again given by Eq.~(\ref{hamiltonianquadraticgeneralAng}). For $\lambda>0$ the final term is now positive. This is plotted in Fig.~\ref{hamiltonianvsradiusrepel}. We see that there is a single extremum, which is a global minimum, and this persists for any choice of particle number $N$. The corresponding solution for the radius in terms of number is
\beq
\R = {a_l + \sqrt{a_l^2+3\,b_{lm}\,c_{lm}\,\n^2}\over b_{lm}\,\n}\,,
\label{Rrepel}
\eeq
where we ignored the negative root, which would correspond to negative radius. This solution is plotted in Fig.~\ref{FigureRadiusNumberRepulsive}. This branch extends to arbitrarily large particle number $N$ for any value of the angular momentum. In the large $N$ regime, the gravitational attraction is balanced by the repulsive self-interaction. This is known as an astrophysical polytrope with equation of state $P\propto\rho^2$. Conversely, in the small $N$ (large $R$) regime, the gravitational attraction is balanced by the repulsive quantum pressure (regardless of the sign of $\lambda$).

In the large $N$ regime, the radius decreases, but asymptotes to a constant, whose value depends on the angular momentum. This value in fact coincides with the minimum value for $R$ reported earlier in the attractive case in Eq.~(\ref{Rmin}), namely 
\beq
\R\to\R_{min}=\sqrt{3\,c_{lm}\over b_{lm}}\,\,\,\,(\mbox{high}\,\,\,\,N)\,.
\eeq
Converting to the physical radius $\R_{90,min}$ that encloses 90\% of the mass, we showed in Eq.~(\ref{R90high}) that this grows as $\R_{90,min}\sim\sqrt{l}/(\ln l)^{1/4}$ in the high $l$ limit.

\section{Astrophysical Implications}\label{Astrophysical}

Here we would like to discuss some possible astrophysical implications of our results. Our emphasis here will be on the QCD axion, whose potential we discussed in Section \ref{Scalars}, though our remarks have relevance for other models too.

By returning to physical variables, we can compute the maximum number of particles $N_{max}$, the maximum mass $M_{max}$, and the minimum clump size $R_{min}$ for the stable (blue) branch as follows
\bea
N_{max} \amp = \amp {\n_{max}\over \ma\sqrt{G\,|\lambda|}} \,\,\,\,\,\,\,\,\,\,\,\, \sim 1.5\times 10^{60}\,{\hat{a}_l\over\sqrt{\hat{b}_{lm}\,\hat{c}_{lm}}}\,(\hat{m}_\phi^{-2}\hat{f}_a\,\hat{\gamma}^{-1/2})\,,\label{NmaxphysicalAng}\\
M_{max} \amp = \amp N_{max}\,\ma \,\,\,\,\,\,\,\,\,\,\,\,\,\,\,\,\sim 2.5 \times 10^{19}\,\mbox{kg}\,{\hat{a}_l\over\sqrt{\hat{b}_{lm}\,\hat{c}_{lm}}}\,(\hat{m}_\phi^{-1}\hat{f}_a\,\hat{\gamma}^{-1/2})\,,\label{MmaxphysicalAng}\\
\rr_{90,min} \amp = \amp {a_l\,(\R_{90}/\R)\over b_{lm}\,N_{max}\,G\,\ma^3} \sim 70\,\mbox{km}\left({\R_{90}\over\R}\right)\sqrt{\hat{c}_{lm}\over\hat{b}_{lm}}\,(\hat{m}^{-1}\hat{f}_a^{-1}\,\hat{\gamma}^{1/2})\label{RminphysicalAng}\,,
\eea
where the constants from the axion potential are normalized to representative values: $\hat{f}_a\equiv f_a/(6\times 10^{11}\,\mbox{GeV})$, $\hat{m}_\phi\equiv \ma/(10^{-5}\,\mbox{eV})$, $\hat\gamma\equiv \gamma/0.3$, and the coefficients are normalized to their zero angular momentum values: $\hat a_l\equiv a_l/a_0$, $\hat b_{lm}\equiv b_{lm}/b_{00}$, $\hat c_{lm}\equiv c_{lm}/c_{00}$. Note that for large angular momentum, the mass and number of particles in the clump increase relatively rapidly as $N_{max}\propto M_{max}\propto l^{3/2}/(\ln l)^{1/4}$, while the size of the clump increases as $R_{90,min} \propto l^{1/2}/(\ln l)^{1/4}$.

As we began discussing in previous papers \cite{Schiappacasse:2017ham,Guth:2014hsa}, it is interesting to compare the maximum number of axions in one of these clumps $N_{max}$ to the expected number of axions that may be somewhat localized due to the random initial conditions of the axion in the early universe. To estimate the latter number, we note that the initial conditions for the axion are essentially laid down by causality. In the scenario in which the PQ symmetry is broken after inflation, the axion is inhomogeneous from one Hubble patch to the next until the axion acquires a mass after the QCD phase transition. This is associated with an initial correlation length $\xi\sim 1/H_{QCD}\sim M_{Pl}/T_{QCD}^2$. The number density can be estimated as $n=\rho/\ma\sim (T_{eq}/T_{QCD})\rho_{QCD}/\ma\sim (T_{eq} T_{QCD}^3)/\ma$, where $T_{eq}$ is the temperature at equality $\sim 0.1$\,eV. So the number of axions within a correlation length can be estimated as \cite{Guth:2014hsa}
\beq
N_\xi\sim {T_{eq}M_{Pl}^3\over T_{QCD}^3\,\ma}\sim 10^{61}\,\hat{m}_\phi^{-1}\,.
\label{Nxi}
\eeq
As we noted in our previous paper \cite{Schiappacasse:2017ham}, for zero angular momentum $l=m=0$, this number $N_\xi$ is $\mathcal{O}(10)$ larger than the maximum number of particles in a clump; given in Eq.~(\ref{NmaxphysicalAng}) for typical axion parameters. So one might be concerned that such conglomerations of particles may struggle to organize into these clump solutions. However, the situation changes when we consider non-zero angular momentum. In this case we find that for $l=|m|\gtrsim5$,  we obtain $N_{max} \gtrsim 10^{61}$, which is now sufficiently large to accommodate all the axions in a typical correlation length. This leads to an interesting possibility that in the early universe, axions may first go into these finite angular momentum states, before perhaps relaxing into the ground state at later times.

\section{Discussion}\label{Conclusions}

This paper has been a natural extension of our previous work on scalar field clumps in Ref.~\cite{Schiappacasse:2017ham} which focussed on spherically symmetric configurations only. Here we generalized this to include arbitrary non-zero angular momentum, which is a form of BEC that extremizes the energy at fixed number and angular momentum. 

We laid out the full set of solutions taking into account gravity and self-interactions. This was self-consistently studied in the non-relativistic regime, which is the regime of most interest as it can lead to long lived configurations, while the highly relativistic regime is short lived due to particle number changing processes. For attractive self-interactions, the solutions organize into a stable and an unstable branch, with a maximum mass that increases with angular momentum; see Fig.~\ref{FigureRadiusNumberAngular}. For repulsive self-interactions, the solutions only contain a single stable branch, without any maximum mass, but with a minimum radius whose value also increases with angular momentum; see Fig.~\ref{FigureRadiusNumberRepulsive}. We found there to be no upper bound on the amount of angular momentum, and that the claim in Ref.~\cite{Davidson:2016uok} of a such a maximum is due to inaccurate numerics.

In our work, we used a convenient ansatz for the spatial profile; namely a modified Gaussian ansatz, which we believe to be rather accurate. Future work is to to compute the properties of these clumps with improved precision, including mode-mode coupling between different spherical harmonics. It would be useful to have a 3-dimensional simulation of these non-spherical clumps. An intriguing possibility is whether there might be interesting behavior if they interact with one another in the galaxy.

While we previously found that the maximum number of axions in a stable BEC is a factor of a few less than the typical number that is expected to appear in a typical correlation length of the axion in the early universe, we found that with non-zero angular momentum, the maximum number of axions in a stable BEC can be larger. This may increase the likelihood that these Bose stars form in the early universe. This may also give rise to new structures in the late universe when axions fall into galaxies and has complicated gravitational interactions, allowing for non-zero angular momentum in local patches. An important subject for future work is to run simulations to determine what percentage of such Bose stars carry large or negligible angular momentum. And furthermore, what percentage of axion dark matter can be in the form of these clumps.

In a forthcoming paper \cite{Hertzberg2018} we will consider the coupling of axions to photons $\Delta\mathcal{L}\sim\phi\,{\bf E}\cdot{\bf B}$ in the context of these Bose stars. We will utilize the results derived here to compute the resonance into photons, including effects from non-zero angular momentum. This may lead to interesting astrophysical signatures.

\section*{Acknowledgments}
We would like to thank McCullen Sandora for useful discussions. MPH is supported by National Science Foundation grant PHY-1720332.

\end{document}